\documentclass[twocolumn,preprintnumbers,amsmath,amssymb,floatfix,superscriptaddress,nofootinbib]{revtex4}

\usepackage{amsmath,hyperref,amssymb}
\usepackage{graphicx}
\usepackage{dcolumn}
\usepackage{bm}
\usepackage{verbatim}
\usepackage{tabularx}
\usepackage{slashed}
\usepackage{cancel}
\usepackage[boxsize=2em,aligntableaux=center]{ytableau}
\usepackage{float}
\usepackage{color}
\usepackage{url}
\usepackage[utf8]{inputenc}

\def\be{\begin{equation}}
\def\ee{\end{equation}}
\def\bea{\begin{eqnarray}}
\def\eea{\end{eqnarray}}

\begin{document}

\vspace*{-30mm}

\title{Constraining primordial black hole abundance with the Galactic 511 keV line}

\author{William DeRocco}
\affiliation{Stanford Institute for Theoretical Physics, \\
Stanford University, Stanford, CA 94305, USA}
\author{Peter W.~Graham}
\affiliation{Stanford Institute for Theoretical Physics, \\
Stanford University, Stanford, CA 94305, USA}

\vspace*{1cm}

\begin{abstract} 

Models in which dark matter consists entirely of primordial black holes (PBHs) with masses around $10^{17}$ g are currently unconstrained. However, if PBHs are a component of the Galactic dark matter density, they will inject a large flux of energetic particles into the Galaxy as they radiate. Positrons produced by these black holes will subsequently propagate throughout the Galaxy and annihilate, contributing to the Galactic 511 keV line. Using measurements of this line by SPI/INTEGRAL as a constraint on PBH positron injection, we place new limits on PBH abundance in the mass range $10^{16} - 10^{17}$ g, ruling out models in which these PBHs constitute the entirety of dark matter.

\end{abstract}

\maketitle

\section{Introduction}

Dark matter (DM) is one of the largest outstanding mysteries of modern physics. A large variety of models relying on new degrees of freedom have been proposed to explain this additional matter content, with masses ranging all the way down to $10^{-22}$ eV~\cite{Hui:2016ltb}. On the high mass end, an explanation that has experienced a resurgence in recent years is that dark matter is composed of primordial black holes (PBHs) that formed as a result of perturbations in the early universe. (See Ref.~\cite{Carr:2016drx} for a recent review of these models and bounds.)

There are a plethora of proposed production mechanisms for PBHs~\cite{GarciaBellido:1996qt,Jedamzik:1999am,Suyama:2006sr,Kohri:2012yw,Kawasaki:2012wr,Bugaev:2013vba,Drees:2011yz,Drees:2011hb,Kawaguchi:2007fz,Kohri:2007qn,Clesse:2015wea,Ballesteros:2017fsr,Ezquiaga:2017fvi,Espinosa:2017sgp,Gross:2018ivp}, but regardless of how the PBHs are produced, any extant PBHs will be producing radiation by means of Hawking evaporation today~\cite{Hawking:1971ei}. The energy spectrum of this radiation is thermal and peaks near the temperature of the black hole, given by 
\be
T_{BH} = \frac{1}{8\pi GM}
\ee
with $M$ the black hole mass. In this paper, we focus mainly on black holes with masses near $10^{16}$ g, which corresponds to a temperature of $\sim 1$ MeV. At these temperatures, the PBHs are sufficiently hot so as to be producing positrons at a large rate.

A prominent effect of PBH DM in this mass range is therefore the injection of large quantities of positrons into the Galaxy. These positrons will eventually annihilate, producing a spectrum of gamma rays peaked at 511 keV. However, measurements of the 511 keV line by the INTEGRAL satellite~\cite{Siegert:2015knp} place a strong limit on the rate at which positrons can be annihilating anywhere in the Galaxy. This in turn constrains the fraction of DM that can be composed of PBHs in this mass range (by the same argument as was used to constrain dark photons in Ref.~\cite{DeRocco:2019njg}).

It should be noted that the contribution of PBHs to the 511 keV line has already been discussed in the literature~\cite{1991ApJ...371..447M, Bambi:2008kx,Fuller:2017uyd}, though mainly only in reference to attempting to explain an excess in the 511 keV line measured in the Galactic center. In Ref.~\cite{1991ApJ...371..447M}, the authors focus exclusively on a scale-invariant distribution and argue that the 511 keV excess could not be produced by PBHs, while in Ref.~\cite{Bambi:2008kx}, the authors find that a log-normal distribution of masses with mean of roughly $6\times10^{16}$ g might yield both the correct 511 keV excess and correct DM abundance. None of the existing literature contains a complete analysis of the constraints on PBH abundance and mass distribution one can infer purely from the measured value of the 511 keV line. In this Letter, we fully determine the precise bounds associated with positron injection by PBHs.

The main existing bounds in this mass range come from observations of both extragalactic~\cite{Carr:2009jm} and Galactic gamma rays~\cite{Carr:2016hva}, as well as recent bounds using measurements of high-energy positrons by the Voyager-1 satellite~\cite{Boudaud:2018hqb}. We demonstrate that positron injection allows us to extend the limits on the fraction of DM that PBHs can comprise by roughly an order of magnitude for PBHs of mass $10^{16}$ to $10^{17}$ g, an otherwise unbounded region of parameter space~\cite{Montero-Camacho:2019jte}.

\section{Positron propagation in the galaxy}
\label{sec: prop}

In this section, we present existing results on positron propagation in the Galaxy and set an upper bound on positron injection energy such that PBH-produced positrons do not escape the Galaxy.

For the PBHs considered in this paper, the production rate of radiation is too low to form an $e^+ e^-$ plasma~\cite{MacGibbon:2007yq}, so radiated positrons will escape the immediate vicinity of the black hole and begin to propagate through the interstellar medium (ISM). Positrons propagating through the Galaxy have a long lifetime in the ISM ($10^5$ to $10^6$ years)~\cite{1993ApJ...405..614C}, during which they diffuse along magnetic field lines and lose energy mainly through Coulomb interactions. Ultimately, after slowing down, they annihilate and can contribute to the 511 keV Galactic flux via either direct annihilation with an electron or the formation of parapositronium~\cite{Alexis:2014rba}.

Though much work has been put into modeling positron propagation in the ISM (see, e.g., Refs.~\cite{2009A&A...508.1099J,galaxies6020039,Alexis:2014rba}), it is a notoriously difficult problem. The distance that a positron propagates depends heavily on the ionization fraction, temperature, and density of the region through which it is passing, as well as the structure of the Galactic magnetic field and the scale of magnetic turbulence. Most simulations of positron propagation have focused on positrons produced within the Galactic plane, as this is where the majority of potential sources lie. The results of these simulations show that positrons with injection energy $E \lesssim 1$ MeV do not propagate much farther than 1 kpc from their source for typical ISM conditions in the plane~\cite{Alexis:2014rba}. 

However, unlike most Galactic positron sources, PBHs are distributed in a spherical halo, so we must  consider the propagation of positrons produced at high latitudes as well. The density of the ISM drops off exponentially above the plane with a scale height of a few hundred parsecs~\cite{1998ApJ...497..759F}. Since the lifetime for positrons propagating through the ISM depends approximately linearly on the density of the ISM (see Eq. 21 of Ref.~\cite{2009A&A...508.1099J}), the propagation distance of positrons increases with increasing height above the Galactic plane.

The Galactic magnetic field is of order 1 $\mu$G~\cite{Jansson_2012}, which results in a gyroradius of order $10^{-11}$ kpc for a 1 MeV positron. As a result, though positrons produced at high latitudes may travel a further distance than those in the plane, they are still tightly confined to magnetic field lines. Their ultimate fate depends upon the large-scale magnetic field structure out of the plane. For realistic models of the poloidal component, such as an A0 dipole~\cite{Han:2004mh,2009IAUS..259..455H} or ``X-shaped field''~\cite{2012ApJ...757...14J,refId1,2014A&A...561A.100F}, the effect of the magnetic field would be to guide roughly half the positrons into the Bulge or Disk. 

In order to place a robust bound, there are two potential situations we wish to avoid: 1) positrons produced in the Galactic plane escaping the Galaxy and 2) positrons produced at high latitudes passing through the Galactic plane and escaping without annihilating. As stated above, simulation results demonstrate that 1 MeV positrons do not propagate far from their source in the Galactic plane, so to avoid these potential complications, we choose to restrict the maximum injection energy of a positron to be 1 MeV. This is very conservative as it is not true that positrons with $E>1$ MeV would necessarily escape the Galaxy without annihilating. For example, if the Galactic magnetic field proves to have a large-scale dipole component as suggested by Han et al.~\cite{Han:2004mh,2009IAUS..259..455H}, positrons that escape the Disk would be funneled back into the Bulge~\cite{Prantzos:2010wi,refId0,Higdon_2009}. Note furthermore that the authors of Ref.~\cite{1991ApJ...371..447M} used a maximum positron energy of 13 (37) MeV for neutral (fully ionized) hydrogen in the Galactic bulge when computing the positron annihilation signal from PBHs, which is significantly less  conservative than the value we have chosen to adopt.

Though we place our final bounds with this conservative restriction, we also display a curve with no upper bound on energy to indicate the potential region that could be bounded by a more careful analysis of positron propagation in the Galaxy (dotted line in Fig.~\ref{Fig: mono}). We have not plotted this all-energy curve for our log-normal bounds, but including all energies improves those bounds in a similar fashion.

\section{Positron production by PBHs}

In this section, we compute the low-energy ($<1$ MeV) positron production rate for all PBHs within our galaxy.

The positron production rate for a PBH of mass $M$ is given by
\be
\frac{d N_e}{dt dE} = \frac{\Gamma_e}{2 \pi}\left[\exp\left(\frac{E}{T_{BH}}\right)+1\right]^{-1}
\ee
with $T_{BH}$ the temperature of the black hole and $\Gamma_e$ the electron absorption probability~\cite{Boudaud:2018hqb}. The absorption probability is a complicated function for energies below $T_{BH}$ that increases by an order of magnitude between the low-temperature and high-temperature limit. The dependence is displayed in Fig. 1 of Ref.~\cite{PhysRevD.41.3052}. In this  paper, we are mainly interested positrons that are produced fairly cold. We therefore must take into account the low-temperature behavior as opposed to simply using the high-energy, geometric optics limit $\Gamma_e = 27 G^2 M^2 E^2$ used frequently in similar analyses.

To determine the total production rate of positrons due to PBHs, we perform the following integral over PBH masses and positron energies to yield $\phi_{\text{inject}}$, the Galactic positron injection rate due to PBH radiation:
\be
\label{eq: rate}
\phi_{\text{inject}} = \int_{E_{\text{min}}}^{E_{\text{max}}} dE \int_{M_{\text{min}}}^{\infty} dM N_{BH}~\eta(M) \frac{d N_e}{dt dE}
\ee
with $N_{BH}$ the number of black holes in the galaxy contributing to the positron flux and $\eta(M)$ the mass distribution function of the PBHs. $E_{\text{min}}$ is taken to be $m_e$ and $E_{\text{max}}$ to be 1 MeV for the reasons described in Section~\ref{sec: prop}. $M_{\text{min}}$ is set at $4\times10^{14}$ g since black holes with lower masses would have lifetimes less than the age of the Universe and would have already evaporated.

To compute $N_{BH}$, we choose to employ a spherically-symmetric Navarro-Frenk-White (NFW) profile~\cite{Navarro_1997} for the galactic DM density
\be
\label{eq: nfw}
\rho(r) = \frac{4 \rho_{\odot}}{\frac{r}{R_{\odot}}(1+\frac{r}{R_{\odot}})^2}
\ee
with the local dark matter density $\rho_{\odot} = 0.4~\mathrm{GeV/cm^3}$ and the distance from the galactic center to Earth $R_{\odot} = 8.2$ kpc~\cite{2017MNRAS.465...76M}. Then the number of PBHs just becomes
\be
N_{BH} = \int dr~\frac{\rho(r)}{M} (4\pi r^2)
\ee
where we take the upper limit of integration to be 15 kpc. For the sake of comparison, we also computed the bounds with a cored Burkert profile~\cite{Nesti:2013uwa} and found that the difference in the resulting bounds was less than $5\%$.

We compute bounds for both a monochromatic PBH mass distribution and a log-normal distribution, a distribution predicted by mechanisms proposed in several recent papers~\cite{Kannike:2017bxn,Bellomo:2017zsr,Calcino:2018mwh}. The log-normal mass distribution, normalized to unity, is defined as
\be
\eta(M) = \frac{1}{\sqrt{2\pi} \sigma M} \exp\left(-\frac{\log^2(M/\mu)}{2\sigma^2}\right)
\ee
with $\mu$ the average mass and $\sigma$ the width of the distribution.

For a monochromatic mass distribution, $\eta(M)$ is simply a delta-function and the expression for $\phi_{\text{inject}}$ becomes even simpler. It is simply
\be
\label{eq: mono}
\phi_{\text{inject}} = N_{BH} \int_{E_{\text{min}}}^{E_{\text{max}}} dE \frac{d N_e}{dt dE}.
\ee

We now have a convenient expression to compute the total positron injection into the galaxy as a function of $\mu$ and $\sigma$ (or $M$ in the monochromatic case). By fixing this injection rate at its upper limit, we can invert the relation to place bounds on the mass spectrum of PBHs in the Galaxy.

\section{INTEGRAL bound}

The current best measurements of the 511 keV line in the Galaxy come from INTEGRAL, a space-based telescope in operation since 2002~\cite{Siegert:2015knp}. INTEGRAL's all-sky map of the 511 keV emission has been used to study potential spatial morphologies for positron annihilation in the Galaxy. INTEGRAL's measurement limits the total production rate of positrons within the Galaxy to less than $\approx2\times10^{43}~e^+~\text{s}^{-1}$ ~\cite{refId01,Prantzos:2010wi}.

The origin of these positrons is as-of-yet not fully understood, though there is a plethora of proposed sources (see Ref.~\cite{Prantzos:2010wi} for a comprehensive review and Ref.~\cite{Fuller:2018ttb} for a more recent proposal). Many of the proposed sources are subject to considerable uncertainty on their production rate and their relative contribution to the total Galactic positron flux. The most relevant sources of uncertainty come from the injection rate by ${}^{56}\text{Ni}$ decays produced in Type Ia supernovae and the injection rate from low-mass X-ray binaries (LMXRBs) since both of these sources can potentially saturate the INTEGRAL bound ($2\times10^{43}~e^+~\text{s}^{-1}$) for optimistic choices of parameters~\cite{Prantzos:2010wi}. Contributions from other sources are in general an order of magnitude smaller than this, e.g. positron production by ${}^{24}\text{Ti}$ decays, which is measured to be $\approx3\times10^{42}~e^+~\text{s}^{-1}$. For a thorough discussion of each potential positron source and its associated uncertainties, see Section IV of Ref.~\cite{Prantzos:2010wi}. 

It is important to note that these astrophysical uncertainties do not affect the bounds presented in this paper, as we do not attempt to add PBH injection on top of existing production mechanisms. Rather, we place a bound by asserting that regardless of other sources and their respective contributions, it is not possible for PBHs to inject twice INTEGRAL's total measured positron production rate into the Galaxy, as this would conflict with observation independent of the contribution of any other sources. One could attempt to place a significantly more stringent bound on PBH positron injection by combining it with other known positron sources in the Galaxy, however due to the aforementioned uncertainties associated with these sources, the resulting bound would not be robust. We have chosen to adopt a very conservative limit in order to avoid this issue.

Since attempting to fully model the propagation of positrons produced by PBHs is beyond the scope of this paper, the most conservative assumption we can make in regards to the spatial morphology of their annihilation would be to assume it matches the best-fitting model of positron annihilation.
By assuming this, the positron annihilation signal from PBHs must exceed the overall flux measured by INTEGRAL to be constrained (instead of simply altering the all-sky pattern of emission), resulting in the weakest possible bounds on PBH abundance.\footnote{It should be noted that we do \textit{not} in general expect the positron annihilation signal from PBHs to have the same spatial morphology as the measured 511 keV line. A PBH signal would likely be less peaked towards the center of the Galaxy than the INTEGRAL measurement since PBHs would be injecting positrons throughout the DM halo, not just in the Galactic bulge. However, in the absence of sufficient data on the large-scale magnetic field structure of the Galaxy, attempting to model the propagation of the positrons and their ultimate annihilation sites is prone to large uncertainty. While the bound that could be placed by modeling propagation would be stronger than the one presented in this paper, it would be less robust.}

Using this assumption, the positron annihilation rate implied by INTEGRAL's measurement of the 511 keV line can be used to limit the rate at which PBHs could be injecting positrons into the Galaxy. Since we wish to restrict ourselves to positrons that will annihilate in the Galaxy without the potential for escape (see Section~\ref{sec: prop}), we place a constraint by conservatively taking the limit on the injection rate of $<1$ MeV positrons by all PBHs in the Galaxy to be $4\times10^{43}~\text{s}^{-1}$ (twice INTEGRAL's measured rate). This allows us to bound PBHs in a straightforward manner. The fraction of DM in the Galaxy that could be composed of PBHs such that this injection rate is not exceeded is simply $f = (4\times10^{43}~\text{s}^{-1})/\phi_{\text{inject}}$, with $\phi_{\text{inject}}$ appropriately restricted in energy. With this, we can readily compute the bounds on PBHs using Eq.~\ref{eq: rate} and Eq.~\ref{eq: mono} for log-normal and monochromatic mass distributions respectively.

Note that the low temperature of the PBHs of interest in this paper means that any potential bound using the in-flight annihilation of positrons such as those employed by Refs.~\cite{Beacom:2005qv} and~\cite{Sizun:2006uh} would be significantly weaker than the bound on the overall 511 keV flux presented here, as these constraints only begin to apply to positrons with injection energies above $3 - 7.5$ MeV.

\section{Results}

\begin{figure}
  \centering
  \includegraphics[width=0.45\textwidth]{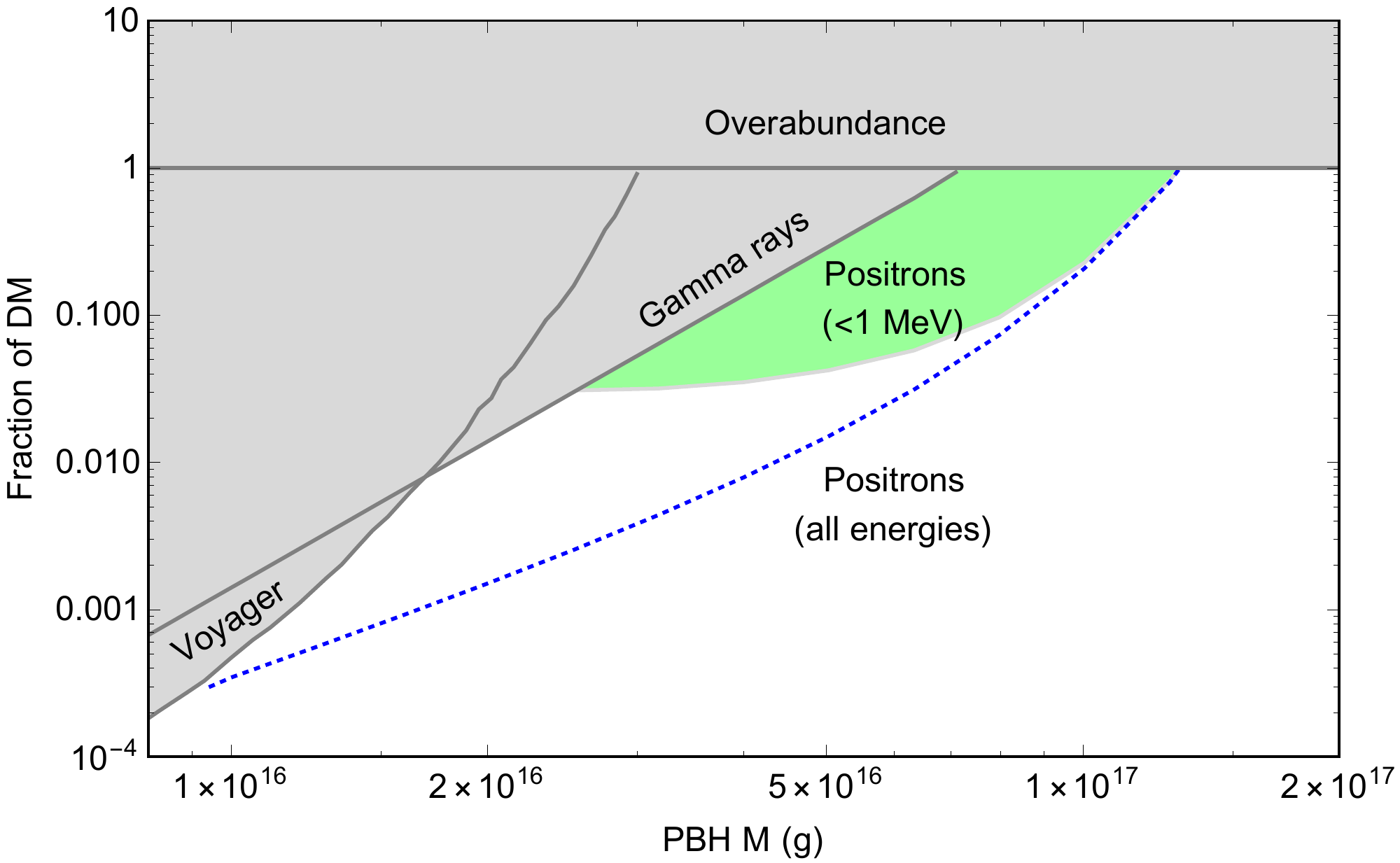}
  \caption{New bounds on PBH DM with a monochromatic mass distribution are shown in green. The blue dotted line is computed in the same manner as the green region but without an upper limit on positron injection energy (see Section~\ref{sec: prop} for a discussion of the 1 MeV limit used to set our final bounds). We display the existing bounds from gamma-ray observations~\cite{Carr:2009jm,Carr:2016hva} and positron measurements by the Voyager spacecraft~\cite{Boudaud:2018hqb} in gray.
    \label{Fig: mono}}
\end{figure}

\begin{figure}
  \centering
  \includegraphics[width=0.45\textwidth]{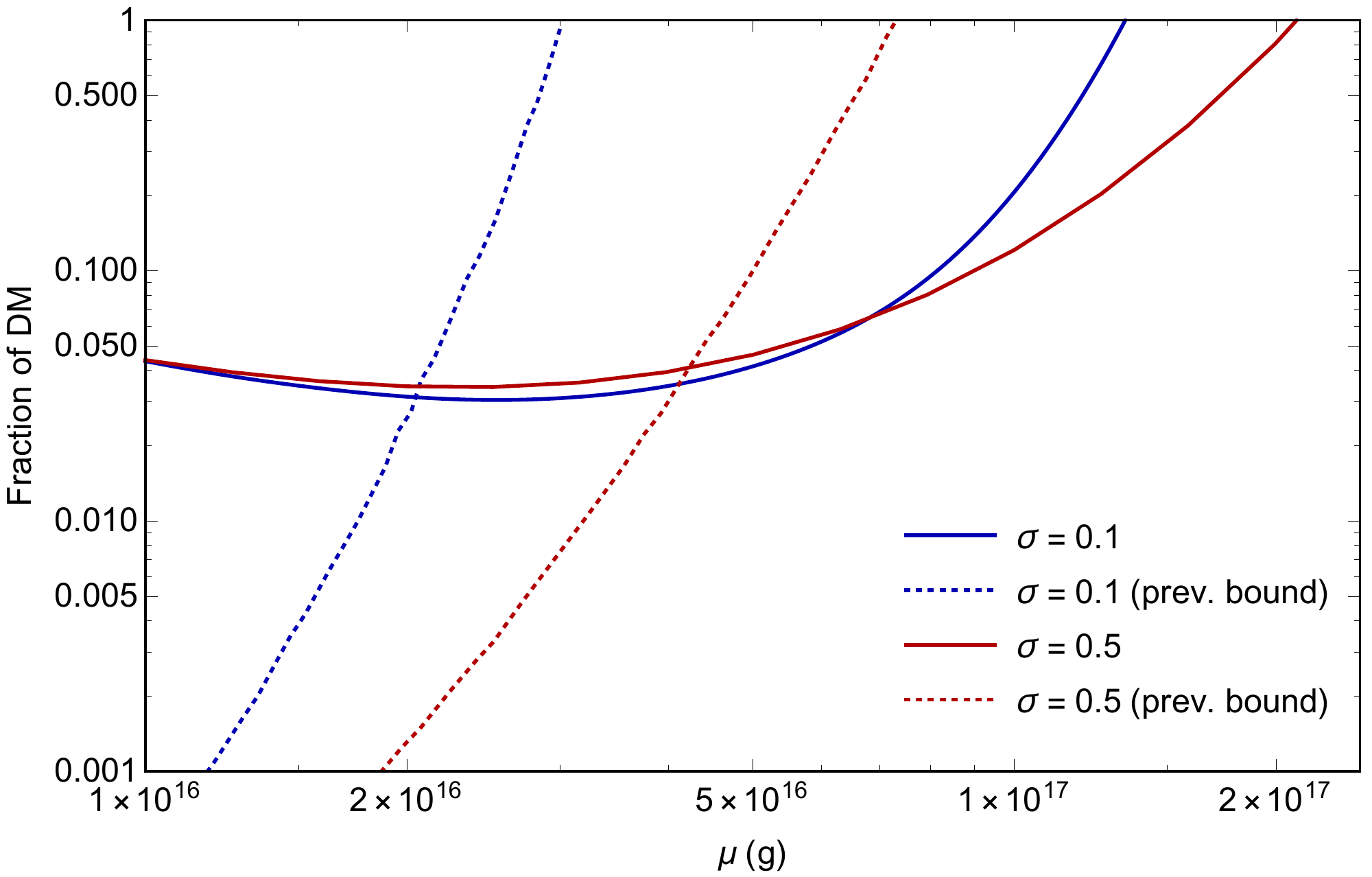}
  \caption{Bounds on PBHs with a log-normal distribution centered at a mass $\mu$ for various widths $\sigma$. The positron injection bounds are shown as solid lines, while the existing Voyager bounds are shown as dotted lines~\cite{Boudaud:2018hqb}.
    \label{Fig: bounds}}
\end{figure}

Our results are displayed in Fig.~\ref{Fig: mono} and Fig.~\ref{Fig: bounds}. In Fig.~\ref{Fig: mono}, we focus exclusively on the case of a monochromatic mass distribution for comparison to existing bounds. The region is bounded above by $f = 1$ since above this line, the abundance of PBHs would exceed the measured density of DM. To lower masses, there is an existing bound from observations of both the extragalactic and Galactic gamma ray flux~\cite{Carr:2009jm,Carr:2016hva} as well as the recent constraint using observations of high-energy positrons by the Voyager spacecraft~\cite{Boudaud:2018hqb}. Note that these bounds are worse than the positron bound discussed in this paper (shown in green) due to the fact that at higher PBH mass, the PBH is cooler and therefore high-energy gamma rays and positrons become Boltzmann-suppressed. In contrast, the positron bound discussed here is based upon low-energy positrons that can annihilate and contribute to the 511 keV line, hence provides stronger constraints for colder (more massive) PBHs.

We have also included a dotted line denoting the region that would be excluded if we did not restrict the positron injection energy to $<$1 MeV. It is clear that towards lower masses where the average energy of emitted positrons is much higher, this would dramatically improve the bounds. As a result, a more careful analysis of positron propagation and annihilation in the Galaxy would have the potential to place significantly stronger constraints.

It is worth additionally remarking that, though attempts to use femtolensing of gamma-ray bursts to place bounds at higher masses have been made, recent reviews have suggested that the assumptions made in these bounds are inappropriate and any bounds disappear when computed carefully~\cite{Katz:2018zrn}. As a result, the limits from these papers are not shown.

In Fig.~\ref{Fig: bounds}, we have plotted the limits for a log-normal mass distribution function with two values of $\sigma$ (0.1 and 0.5 in blue and red respectively). It is clear that for these values of $\sigma$, the positron bound places stronger constraints than existing bounds (taken from Ref.~\cite{Boudaud:2018hqb}) at the high-mass end. However, as the distribution becomes wider, the high-energy tail of positrons improves the Voyager bounds while not significantly affecting the 511 keV line bounds. As such, beyond $\sigma \approx 1$, the Voyager bounds are more restrictive than our positron injection bound for all masses.

\section{Conclusion}

Using the positron annihilation rate implied by INTEGRAL's measurements of the Galactic 511 keV line, we have placed strong constraints on the fraction of dark matter that can be comprised of primordial black holes in the mass range $10^{16} - 10^{17}$ g. We show that such PBHs cannot comprise the entirety of dark matter and improve constraints on both monochromatic and log-normal mass distributions by roughly an order of magnitude in this mass range. Future measurements of the Galactic magnetic field structure and distribution of ISM subcomponents would allow the conservative assumptions used in this paper to be relaxed, potentially resulting in even stronger constraints.

\section*{Acknowledgements}

The authors would like to express their gratitude for the support provided by DOE Grant DE-SC0012012, by NSF Grant PHY-1720397, the Heising-Simons Foundation Grants 2015-037 and 2018-0765, DOE HEP QuantISED award \#100495, and the Gordon and Betty Moore Foundation Grant GBMF7946.

\bibliography{ref}

\end{document}